\documentstyle[12pt]{article}
\def\be{\begin{equation}}
\def\ee{\end{equation}}
\begin{document}
\title{\bf Nuclear Matter Properties in Derivative Coupling Models 
 Beyond Mean-Field Approximation}
\vskip .3cm
\author{A. Delfino $^{1}$ \thanks{Partially supported by CNPq of 
Brasil},
 M. Malheiro $^{2}$ \thanks{Partially
supported by CAPES of Brasil, Permanent address:
Instituto de 
F\'\i sica, Universidade Federal Fluminense, Niter\'oi,
R. J., Brasil}
and D. P. Menezes $^{3}$ 
\thanks{Partially supported by CNPq of Brasil, email address: 
FSC1DPM@FSC.UFSC.BR }\\ \\ $^{1}$ Instituto de 
F\'\i sica, Universidade Federal Fluminense, \\ 24210-340, Niter\'oi,
R. J., Brasil
\\
${^2}$ Department of 
Physics, University of Maryland, \\ College Park, Maryland 20742-4111, USA
\\
${^3}$ Depto de F\'{\i}sica - Universidade Federal de Santa Catarina
 \\  C.P. 476  - 88.040-900 - Florian\'opolis - SC - Brazil}
\vspace{.3 cm} 
\date{\today}
\maketitle
{\it Abstract} :
The structure of infinite nuclear matter is studied with two of the Zimanyi - 
Moszkowski (ZM)
models in the framework of a relativistic approximation which takes into 
account Hartree
terms and beyond and is compared
with the results which come out of the relativistic Hartree - Fock 
approach in the linear Walecka model. The simple treatment applied to
these models can be used in substitution to the more complicated
Dirac - Brueckner - Hartree - Fock method to perform future calculations in finite nuclei.
\\
PACS  \# 21.65.+f, 12.40.-y, 21.60.Jz

\newpage
Conventional many- body calculations which do not consider mesonic degrees of
freedom are not reliable for the study of nuclear matter where high densities
are present. For these purposes, a relativistic calculation must be performed. 
Quantum chromodynamics (QCD) is the fundamental theory of the strong
interaction and hence it should explain possible modifications of
hadron properties in the nuclear medium. However, typical nuclear
phenomena at intermediate and low energies cannot be analitically derived
from QCD although one hopes that QCD will be solved numerically on the
lattice in a near future. Meanwhile we are left with the construction of
phenomenological models in order to try to describe nuclear phenomena
and bulk properties.
Walecka and collaborators used such a kind of theory for the first time around  
1974 to describe the nucleon - nucleon interaction \cite{wc}. Since then
the Walecka model \cite{sw} has been widely used to describe the properties of
nuclear matter as well as some properties of finite nuclei. This model is based on a phenomenological treatment of the
hadronic degrees of freedom and, because of this fact, it is also known as
QHD-I (quantum hadrodynamics), which consists in a renormalizable relativistic 
quantum field theory.
 The early version of this model considers a
scalar ($\sigma$) meson field and a vector ($\omega$) meson field coupled to the
baryonic (nucleon) field. Besides the relativistic mean field calculation, the
Walecka model has also been used in a more complete treatment, the relativistic
Hartree- Fock approximation \cite{sw}, \cite{rhf}, \cite{fra}.

Some of the drawbacks of the model are that the effective nucleon mass
obtained at high densities is too small
and its incompressibility at the energy density saturation is too large. 
To avoid this problem, Zimanyi and Moszkowski \cite{zm} introduced an 
alternative coupling between the scalar meson and the nucleon 
(which is a scalar derivative coupling) and another coupling between the 
scalar and vector mesons. We should stress that a microscopic foundation
for derivative scalar coupling models has been derived from the 
relativistic $SU(6)$ model \cite{myazaki}.
It has already been investigated, in a mean- field approximation
\cite{dcm},\cite{dcmb}, how this derivative coupling Zimanyi - Moszkowski (ZM) model differs
from the usual Walecka 
model when the effective nucleon mass, the energy density, the incompressibility and 
other important features are calculated. 

To try to solve the above mentioned drawbacks of the Walecka model,
more sophisticated treatments have also been developed. One of them 
\cite{bernar} refers to a relativistic  Hartree - 
Fock approximation in a non - linear Walecka model for nuclear matter and 
finite nuclei which considers $\sigma$, $\omega$, $\pi$ and $\rho$ mesons.
In this work, the $\sigma$ - meson mass is density dependent and the non
linear character of its related Hamiltonian makes the calculation of the
exchange contributions to the energy rather complicated, which requires
an approximation for the calculation of this Fock - like terms. Even
though, the inclusion of the exchange terms and the isovector mesons
($\pi$, $\rho$) turn out to be important for the description of the
energy and the incompressibility in nuclear matter and of the spin -orbit
interaction. Another successful treatment utilizes the so called
Relativistic Dirac-Brueckner-Hartree-Fock method (RDBHF
) \cite{dbhf}. In this case
the coupling constants for the sigma and omega mesons become density
dependent and they actually decrease with increasing energy. The 
computational effort also increases considerably since self - consistency
for the self energies is obtained by considering different coupling 
constants at each density. Nevertheless this sort of calculation has 
proved very good when applied to finite nuclei \cite{findbhf} and the
authors concluded that
Fock terms are small but important and should not be neglected.

It is important to stress that the ZM models also belong to the category of descriptions involving density-dependent coupling constants
\cite{dcmb}. The description underlying the approach known as relativistic density-dependent Hartree-Fock \cite{ddhf} reproduces finite nuclei and nuclear matter saturation properties using coupling constants that are fitted, at each density value, to the RDBHF self-energy terms. The good agreement obtained for the ground state properties of spherical nuclei lends support to this sort of  description involving density-dependent coupling constants.

In the present work, we propose an alternative way of considering in a very simple manner direct and exchange terms in the infinite nuclear matter
to obtain the properties of the
ZM models mentioned above and compare them with the results which come out 
of the relativistic Hartree - Fock approximation for the
Walecka model. Our aim is to use these derivative scalar coupling models
to obtain bulk properties in finite nuclei in a forthcoming work and 
for this purpose, based on the conclusions drawn in refs. \cite{fra},
\cite{bernar}, \cite{findbhf}, we assume that the exchange terms will be necessary. In
this sense, the present work has also a didatic meaning.

In the recent literature three different possibilities have been
considered for the coupling of the nucleon with the mesons in the
Zimanyi - Moszkowski (ZM) model (for a review, check ref. \cite{dcm}). In
this letter we consider two of them. The first one is known as the original ZM model and the second one has been chosen because is the one
which gives better results for the nuclear matter in the mean field approximation 
and we call it ZM3 (to be consistent with the definition in other
published material). To start with, we write the Lagrangian density 
for the ZM  and ZM3 models respectively as \cite{zm}:

$${\cal L}_{zm}=-{\bar \psi} M \psi + 
(m^*_{zm})^{-1}{\bar \psi}(i \gamma_{\mu} \partial^{\mu} 
- g_v \gamma_{\mu} V^{\mu}) \psi $$
\be
+{1\over 2}(\partial_{\mu} \phi \partial^{\mu} \phi - m_s^2 \phi^2)
-{1\over 4} F_{\mu \nu} F^{\mu \nu} + {1\over2} m_v^2  V_{\mu} V^{\mu},
\label{e1}
\ee
and
$${\cal L}_{zm3}=-{\bar \psi} M \psi + 
(m^*_{zm})^{-1} {\bar \psi} i \gamma_{\mu} \partial^{\mu} \psi 
- g_v {\bar \psi} \gamma_{\mu} V^{\mu} \psi $$
\be
+{1\over 2}(\partial_{\mu} \phi \partial^{\mu} \phi - m_s^2 \phi^2)
-{1\over 4} F_{\mu \nu} F^{\mu \nu} + {1\over2} m_v^2  V_{\mu} V^{\mu},
\label{e2}
\ee
where 
\be
m^*_{zm}=(1+ {g_s \phi \over M})^{-1} \label{m*}
\ee
and $\psi$, $\phi$ and $V^{\mu}$ are field amplitudes 
for the nucleon, the scalar-isoscalar meson, 
the vector-isoscalar meson, 
$F^{\mu \nu} = \partial^{\mu} V^{\nu}-\partial^{\nu} V^{\mu}$ and 
$M$ is the bare nucleon mass. Notice that these derivative coupled 
Lagrangians are Lorentz invariant, but they are not renormalizable.
The physical meaning of this modified couplings is that the kinetic
fermionic term describes the motion of a particle with
an effective mass $M^*$ instead of the bare mass $M$ present in the
conventional Walecka model.

In ref. \cite{zm}, a rescaled Lagrangian is obtained
from the above equations, with the rescaling of the fermion wave function as
$\psi \rightarrow \sqrt{m^*_{zm}} \psi$ in eqs.(\ref{e1}) and (\ref{e2})
and the rescaling of the field $V_{\mu}$ as 
$V_{\mu} \rightarrow m^*_{zm} V_{\mu}$ in eq.(\ref{e2})
and it reads
$$
{\cal L}_R={\bar \psi}(i \gamma_{\mu} \partial^{\mu} - M + 
{m^*_{zm}}^{\beta} g_s \phi) \psi$$
$$+ {m^*_{zm}}^{\alpha} (-g_v {\bar \psi} \gamma_{\mu} V^{\mu} \psi
-{1\over 4} F_{\mu \nu} F^{\mu \nu} + {1\over2} m_v^2  V_{\mu} V^{\mu})
$$
\be
+{1\over 2}(\partial_{\mu} \phi \partial^{\mu} \phi - m_s^2 \phi^2),
\label{er}
\ee
where, for the Walecka model $\alpha=0$, $\beta=0$, for ZM $\alpha=0$, 
$\beta=1$ and for ZM3 $\alpha=2$, $\beta=1$.

In a very simple description of the ZM and ZM3 models they can be viewed
as models with non-linear effective scalar coupling constants $g_s^*$ and
$g_v^*$ in such a way that the new Lagrangians and related quantities
can be obtained simply by substituting the old coupling constants in the
Walecka model by the new ones \cite{dcmb},i.e.:
$${\cal L}_{zm}= {\cal L}_{Walecka}(g_s \rightarrow g_s^*)$$
and
$${\cal L}_{zm3}= {\cal L}_{Walecka}(g_s \rightarrow g_s^* , g_v \rightarrow 
g_v^*),$$
where $g_s^*=m^*_{zm} \cdot g_s$ and $g_v^*=m^*_{zm} \cdot g_v$.
Instead of the above substitutions, which can be cumbersome to be carried
out in the self energy expressions, we utilize a new subscription which
introduces modifications in the meson masses only \cite{delfino}:  
\be
{\cal L}_{zm}= {\cal L}_{Walecka}(m_s \rightarrow m_s^*) \label{zm}
\ee
and
\be
{\cal L}_{zm3}= {\cal L}_{Walecka}(m_s \rightarrow m_s^* , m_v \rightarrow 
m_v^*), \label{zm3}
\ee
where
$m_s^*=m_s / m^*_w$ and $m_v^*=m_v / m^*_w$, where, in this case,
\be
m^*_w=1- {g_s \phi \over M}. \label{nm*}
\ee

In the Walecka model the relativistic Hartree-Fock equations 
\cite{rhf},\cite{fra} are obtained by using Dyson's
equation to sum to all orders the self-consistent tadpole and exchange
contributions to the baryon propagator
\begin{equation}
G(k)=G^0(k)+G^0(k) \Sigma(k)G(k),
\label{e5}
\end{equation}
\noindent
where $\Sigma$ is the proper self-energy. 
Because of translational and rotational invariances in the
rest frame of infinite nuclear matter and the assumed invariance under
parity and time reversal, the self-energy may be written
as \cite{rhf},\cite{sha}
\be
\Sigma(k)=\Sigma^s(k)-\gamma_0 \Sigma^0(k)+
\vec \gamma \cdot \vec k \Sigma^v(k).
\label{e7}
\ee
 
Coupled integral equations are then solved for the self-energies in the
so-called Dirac-Hartree-Fock approximation \cite{rhf}. In this approximation
just the contributions from real nucleons in the Fermi sea are kept 
in the baryon propagators.
The effects of virtual nucleons and anti-nucleons on the medium
are neglected as usually done in the literature
\cite{sw}, \cite{rhf}.
The nucleon propagator in a Fermi sea with
Fermi momentum $k_F$ is then written
as (the nuclear density is $\rho_B = 2 k_F^3/{3 \pi^2}$)
\begin{equation}
G(k)=(\gamma_{\mu} {k^{\mu}}^* + M^*(k)){\pi i\over E^*(k)}
\delta(k^0 - E(k)) \theta(k_F- \vert \vec k \vert ), \label{e8}
\end{equation}
where
\begin{equation}
{k^{\mu}}^*=k^{\mu}+\Sigma^{\mu}(k)=\left( k^0+\Sigma^0(k),
\;\vec k(1 + \Sigma^v(k))\right) \;,
\end{equation}
\begin{equation}
E^*(k)=\sqrt{(\vec k^*)^2 + {M^*(k)}^2} \label{e10}
\ee
\be
M^*(k)=M + \Sigma^s(k), \label{e11}
\end{equation}
\noindent
and $E(k)$ is the single-particle energy, which is the solution of the
transcendental equation
\begin{equation}
E(k) = [E^*(k) - \Sigma^0(k)]_{k^0 = E(k)}\;.
\label{ek}
\end{equation} 
Performing the $q^0$ and angular integrals in the expressions for the
various components $\Sigma^s$, $\Sigma^0$, $\Sigma^v$ of the self-energy,
three coupled nonlinear integral equations are then obtained. For the 
Walecka model, the equations for the self - energies are
given in ref. \cite{sw}, (pg. 131) and we do not reproduce them here.
In the ZM model the very same steps are performed in order to obtain 
these three components of the self energy, with the exception that we 
have included an approximation. This approximation amounts to
considering $m^*$, defined in eq. (\ref{nm*}), as a function only of the 
momentum, i.e.,
\be 
m^*_w(k)={M^*(k) \over M} \label{e13}.
\ee

Because of this fact the Hartree terms are exactly calculated, but the 
Fock ones are somewhat approximated as in ref. \cite{bernar}.
 The exact calculation would imply in
considering $m^*$ as a function of the field operator $\phi$ and this
is complicated. However,we still have an improvement on the Hartree 
calculation with the introduction of the exchange terms. 
The nonlinear integral equations for the ZM3 model are the 
same expressions as obtained in the Walecka model with the following
modifications:

1.) All $m_s$ are substituted by $m_s^*$ and all $m_v$ are substituted
by $m_v^*$;

2.) As a consequence of 1.), the functions $\Theta_i(k,q)$ and $\Phi_i(k,q)$
are modified because of their dependence on $A_i(k,q)$, which becomes  
\begin{equation}
A_{i}(k,q)= {\vec k}^2 + {\vec q}^2 + {m_i^*}^2 - [E(q) - E(k)]^2\;,
\end{equation}
\noindent
where $i=\sigma$ or $v$.

For the ZM model just the modifications related to the mass of the sigma meson
are carried out.
All self-energies are evaluated at the self-consistent single-particle
energies, $q^0 = E(q)$ and the equations for $\Sigma^s$, $\Sigma^0$ and
$\Sigma^v$ are solved by a direct iteration procedure.
  
We choose to normalize the model parameters using the bulk binding energy 
and
saturation density of nuclear matter as usual. As normally done in
calculations with the Walecka model, we identify
the vector meson with the $\omega$ whose mass is $m_v = 783$ MeV
and set $m_s= 550$ MeV for the scalar meson mass. For the nucleon mass we
take $M = 939$ MeV. The energy density for the ZM3 model once the
substitution given in eq. (\ref{zm3}) is performed yields:
$$
{\cal E} = {2\over \pi^2} \int^{k_F}_0 k^2 E(k) dk
- {g_v^2 \over 2 {m_v^*(k_F)}^2} \rho_B^2 +$$
\be
{2 g_s^2 \over {m_s^*(k_F)}^2 \pi^4} 
\left( \int^{k_F}_0 k^2 {M^*(q)\over E^*(q)} dk \right)^2 
+ 2 g_s^2 I_{\sigma} + 4 g_v^2 I_v ,
\label{dens}
\ee
\noindent
where the ${I_i}$, $i=\sigma, v$ are integrals of the following form
\be
I_i={1\over (2 \pi)^6}
\int^{k_F}_0 {d^3 k \over E^*(k)}
\int^{k_F}_0 {d^3 q \over E^*(q)}
D^0_i(k-q) F_i(k,q) H_i(k,q) 
\label{ifg}
\ee
\noindent
with the functions $F_i$ and $H_i$ given by
\be
F_i(k,q) = 1/2 - (E(k)-E(q))^2 D^0_i(k-q),
\label{ffp}
\ee
\begin{eqnarray}
&&H_{\sigma}(k,q)=k^*_{\mu} q^{* \mu} + M^*(k)M^*(q),
\nonumber \\
&&H_v(k,q)=k^*_{\mu} q^{* \mu} -2 M^*(k)M^*(q);,
\end{eqnarray}
\noindent
and the ${D_i}'s$ are the meson propagators
\begin{equation}
D_i^0(k) = {1\over k_{\mu}^2-{m_i^*}^2 + i\epsilon}\;.
\end{equation}

To saturate the binding energy per nucleon at $-16.15$ MeV
at the Fermi momentum of $1.14 fm^{-1}$ we use
$g_s^2 = 100$ and $g_v^2= 60$ for the ZM model. In the ZM3
model, the energy saturates at $-15.18$ MeV at $1.38 fm^{-1}$
for $g_s^2=102$ and $g_v^2=105$. In the Walecka model, the
energy saturates at $-15.75$ MeV at  $1.31 fm^{-1}$
for $g_s^2 = 108.8$ and $g_v^2= 149.4$.

We have then plotted the effective 
nucleon mass $M^*/M$ in function of $k_F$ for the three models in fig. 1. 
One of the reasons for the introduction of the derivative coupling used in 
the ZM and ZM3 models is the small value of the effective nucleon mass at saturation density obtained 
with the Walecka model. We can observe that this problem is
corrected in the new models. $\Sigma^0$ and $\Sigma^v$ in terms of the
momentum also decrease much less in the ZM and ZM3 models than in the
Walecka model. 

Other important quantities to be analyzed are the scalar and vector 
potentials and
the incompressibility. To derive expressions for the potentials, 
we start from the Dirac equation:
\be
\left((1+\Sigma^v)\vec \alpha \cdot \vec k + \gamma_0(M+ \Sigma^s) -
\Sigma^0 \right) \psi= E \psi,
\ee
and after some simple algebraic manipulations, it can be rewritten as
\be
\left( \vec \alpha \cdot \vec k + \gamma_0(M+ 
{\Sigma^s - M \Sigma^v \over{1+ \Sigma^v}}) 
+{ - \Sigma^0 + E \Sigma^v \over{1 + \Sigma^v}} \right) \psi= E \psi,
\ee
from where we may define in a natural way
\be
S={\Sigma^s - M \Sigma^v \over{1+ \Sigma^v}}
\ee
as the scalar potential and
\be
V={ - \Sigma^0 + E \Sigma^v \over{1 + \Sigma^v}}. 
\ee
as the vector potential, where E given by eq.(\ref{ek})
can be written as $E= {\cal E}/\rho$ at the saturation density. In table 1 these
potentials are displayed for the Walecka and the ZM models in the
mean field and Hartree - Fock calculations. 
The $V-S$ quantity is related to the spin- orbit splitting in finite
nuclei and the $V+S$ quantity corresponds to the real part of the 
optical potential for the zero three-momentum. 
As there is some evidence for the strong potentials
of the Walecka model, we believe that the ZM3 model can probably 
provide good results for finite nuclei while the ZM potentials are
too weak.

Concerning the incompressibility $K$, it has also been calculated 
for the three models, where
\be
K=9 \rho_0^2 {\partial^2 \over \partial \rho^2} {{\cal E}\over \rho} 
\vert_{\rho=\rho_0}
\ee
with $\rho_0(k)=\rho_B (k/k_F)^3$. The results are also shown in table 1. 
According to
ref \cite{blaizot}, the expected incompressibility value is 
$K=210 \pm 30$ MeV,
which means that the results obtained for the ZM  and ZM3 models 
are of great improvement in
comparison with the results coming from the Walecka model.

To conclude, we would like to comment that we have calculated some
relativistic features which are important for the understanding of
nuclear matter within the context of the ZM and ZM3 models. For the present
calculation we have taken into account the Hartree terms (related to
the direct diagrams for the baryon propagator) and Fock - like terms
(related to the exchange diagrams) with the help of the approximation
given by eq.(\ref{e13}). We have calculated
the effective mass, the incompressibility and
the scalar and vector potentials and obtained good results out of the
ZM3 model. The same treatment applied to these models can be extented
to finite nuclei calculation in substitution to the more complicated
relativistic Dirac - Brueckner - Hartree - Fock  approach.
This investigation is under way.

\newpage
\begin{center}
{\bf Figure Captions}
\end{center}

Fig.1) Effective nucleon mass as a function of $k_F$. The solid line
stands for the Walecka model, the dashed line for the ZM model and the
dot - dashed line for the ZM3 model.
 
\begin{center}
{\bf Table Caption}
\end{center}

Table 1) Scalar, vector potentials and incompressibility at nuclear matter saturation density are shown for the
Walecka and the ZM models. MF stands for mean-field and HF for Hartree-Fock approximation.

\newpage
\centering
\begin{tabular}[pos]{|c|c|c|c|c|c|} \hline
 models &  S  (MeV)  &  V  (MeV)  &  V + S (MeV) & V - S  (MeV) 
&  K  (MeV)  \\ \hline
 Walecka - MF  & -431.02  & 354.12  
& -76.87 & 785.18 & 550.82 \\ \hline
 Walecka - HF  & -458.31  & 379.01  
& -79.30 & 837.32 & 585.00 \\ \hline
 ZM - MF  & -140.64 & 82.50  
& -58.13 & 223.13 & 224.71 \\ \hline
 ZM - HF & -177.5 & 109.24 
& -68.26 & 286.74 & 298.47 \\ \hline
 ZM3 - MF & -267.00 & 203.71 
& -63.28 & 470.71 & 155.74 \\ \hline
 ZM3 - HF  & -264.94 & 178.92 
& -86.02 & 443.86 & 161.81 \\ \hline
\end{tabular} 

\newpage
\begin{thebibliography}{99}

\bibitem{wc}J.D. Walecka, Ann. Phys. 83 (1974) 491, S.A. Chin and J.D. 
Walecka, Phys. Lett. 52 B (1974) 24.

\bibitem{sw}B.D. Serot and J.D. Walecka, Adv. Nucl. Phys. 16 (1985) 1.

\bibitem{rhf} C.J. Horowitz and B.D. Serot, Nucl. Phys. A399 (1983) 529;
T. Matsui and B.D. Serot, Ann. Phys. 144 (1982) 107;
G. Krein, D.P. Menezes and M. Nielsen, Phys. Lett. B 294 (1992) 7.

\bibitem{fra} A. Bouyssy, J.-F. Mathiot, N.V. Giai and S. Marcos, Phys.
Rev. C 36(1987)380.

\bibitem{zm} J. Zimanyi and S.A. Moszkowski, Phys. Rev. C 42 (1990) 1416.

\bibitem{myazaki} K. Myazaki, Prog. of Theo. Phys. 93 (1995) 137.


\bibitem{dcm} A. Delfino, C.T. Coelho and M. Malheiro, Phys. Lett. B 345
(1995) 361; A. Delfino, C.T. Coelho and M. Malheiro, Phys. Rev. C 51 (1995) 
2188

\bibitem{dcmb} A. Delfino, M. Chiapparini, M. Malheiro, L.V. Belvedere and
 A. Gattone, nucl-th/9602004 to appear in Z. Phys. A.

\bibitem{bernar} P. Bernardos,V.N. Fomenko, N.v. Giai, M.L. Quelle,
S. Marcos, R. Niembro, L.N. Savushkin, Phys. Rev. C 48 (1993) 2665.

\bibitem{dbhf} M.R. Anastasio, L.S. Celenza, W.S. Pong and C.M. Shakin,
Phys. Rep. 100 (1978) 327; R. Brockmann and R. Machleidt, Phys. Lett. 
B 149 (1984) 283; B. ter Haar and R. Malfliet, Phys. Rep. 149 (1987) 207;
C.J. Horowitz and B.D. Serot, Phys. Lett. B 137 (1984) 287; Nucl. Phys A
464 (1987) 613.

\bibitem{findbhf} R. Fritz and H. Muther, Phys. Rev. Lett. 71 (1993) 46. 

\bibitem{ddhf} Hua-lin Shi, Bao-qiu Chen and Zhong-yu Ma, Phys. Rev. C 52
 (1995) 144.

\bibitem{sha}L.S. Celenza and C.M. Shakin, Relativistic nuclear physics:
Theories of structure and scattering (World Scientific, 1986).

\bibitem{delfino} A. Delfino, preprint IFUFF, in preparation 

\bibitem{blaizot} J.P. Blaizot, Phys. Rep. C 64 (1980) 171.

\end{thebibliography}
\end{document}